# Mendeley readership as a filtering tool to identify highly cited publications[1]

*Zohreh Zahedi, Rodrigo Costas and Paul Wouters*

*z.zahedi.2@cwts.leidenuniv.nl; rcostas@cwts.leidenuniv.nl; p.f.wouters@cwts.leidenuniv.nl*

*CWTS, Leiden University, P.O. Box 905, Leiden, 2300 AX (The Netherlands)*

**Abstract**

This study presents a large scale analysis of the distribution and presence of Mendeley readership scores over time and across disciplines. We study whether Mendeley readership scores (RS) can identify highly cited publications more effectively than journal citation scores (JCS). Web of Science (WoS) publications with DOIs published during the period 2004-2013 and across 5 major scientific fields have been analyzed. The main result of this study shows that readership scores are more effective (in terms of precision/recall values) than journal citation scores to identify highly cited publications across all fields of science and publication years. The findings also show that 86.5% of all the publications are covered by Mendeley and have at least one reader. Also the share of publications with Mendeley readership scores is increasing from 84% in 2004 to 89% in 2009, and decreasing from 88% in 2010 to 82% in 2013. However, it is noted that publications from 2010 onwards exhibit on average a higher density of readership vs. citation scores. This indicates that compared to citation scores, readership scores are more prevalent for recent publications and hence they could work as an early indicator of research impact. These findings highlight the potential and value of Mendeley as a tool for scientometric purposes and particularly as a relevant tool to identify highly cited publications.

**Introduction and background**

Scholars use social media tools for different purposes, for example to collaboratively distribute scientific information, share knowledge and ideas, and communicate with their peers (Gruzd, Staves, & Wilk, 2012). Among the different altmetric sources, Mendeley is one of the most important online reference managers with more than 4 million users worldwide[1], and is especially popular among students and postdocs (Zahedi, Costas, & Wouters, 2014b; Haustein & Larivière, 2014). Mendeley exhibits a high coverage of scientific publications, with coverage values higher than 60% or even 80% for WoS publications depending on the field (Costas, Zahedi, & Wouters, 2015b).

*Meaning of Mendeley readership*

Mendeley collects usage statistics per document as they are added by the different users to their private libraries. These statistics are commonly known as "readership statistics", although in reality the metrics don't necessarily reflect the actual 'reading activity' by Mendeley users. For example, scholars do not necessarily always 'read' the scholarly outputs that they save in Mendeley (Mohammadi, Thelwall, & Kousha, 2015). Thus the actual meaning of "readership" in Mendeley is not fully known yet and this introduces a conceptual constraint on the actual value that the act of "saving" a document in Mendeley may have. Moreover, not all scholars are familiar with Mendeley; instead they may use other reference management tools in their scholarly process of reading and referencing papers (or none at all). Therefore, the usefulness of Mendeley readership strongly depends on the coverage and presence of users from different disciplines, countries, academic statuses, ages, etc. Another important issue is that Mendeley does not provide any information on the timestamp (date) when a given document has been added by a user to her/his library[2]. Therefore, important information on the patterns of readership scores accumulation over time for the saved publications is still lacking, making the adequate study of readership history patterns impossible.

---

[1] This is a preprint of an article accepted for publication in *Journal of the Association for Information Science and Technology* copyright © 2017 (Association for Information Science and Technology), DOI: 10.1002/asi.23883.

*Characteristics of Mendeley as a scientometric tool*

Previous studies have shown moderate correlations between readership and citation scores (see Zahedi, Costas, & Wouters, 2014a; Haustein et al., 2014b; Thelwall & Wilson, 2015). The correlations between Mendeley readership and citation scores are higher than the correlations between citations and other altmetric indicators (Thelwall et al, 2013; Costas, Zahedi, & Wouters, 2015a), thus a stronger similarity between these two metrics in comparison to other altmetric sources can be assumed. Furthermore, publications with more Mendeley readership scores tend to have higher number of citations and are published in journals of higher impact compared to those with less or without any readership (Zahedi, Costas, & Wouters, 2014a). All these results suggest that Mendeley can be a relevant tool for scientometric purposes, and for example, suggestions of normalization of the number of readership by discipline have already been proposed (Haunschild & Bornmann, 2016).

Some other important features of Mendeley are that these readership statistics include data about the academic's status, disciplines and countries of the Mendeley users. This information on the academic's disciplinary and geographic background of the different users helps to better understand the saving patterns of scientific publications by different groups of users (Haunschild, & Bornmann, 2015; Haunschild, Bornmann & Leydesdorff, 2015; Thelwall & Maflahi, 2015). Another important characteristic of this tool is that readership data tend to be collected and made available before citation is recorded by any citation database. Thus, Mendeley readership scores can be seen as evidence of 'early' impact of scientific publications (Maflahi & Thelwall, 2016). However, as mentioned before, due to the lack of historical information reported by Mendeley regarding the date and time at which readership happened, it is not possible, at this time, to perform reliable analyses regarding the prediction of future citations using Mendeley readership scores.

*Identification of highly cited publications*

Studying highly cited publications and the factors influencing them is an important topic in the scientometric literature (Ivanović & Ho, 2014; Aksnes, 2003). Although being highly cited does not always truly reflect the higher research quality of publications (Waltman, Van Eck, & Wouters, 2013), high citedness can be a characteristic sign of relevant or even potential 'breakthrough' papers (Schneider & Costas, 2014) as well as an indicator of scientific excellence (Bornmann, 2014) and the share of such highly cited papers is considered as a relevant indicator in research evaluation in a large number of fields (Abramo et al., 2015; Tijssen et al., 2002). Therefore the identification of highly cited publications can be considered as a critical element in bibliometric research as well as research evaluation.

The use of journal level impact indicator in order to capture the "quality" of individual scientific publications has been widely criticized in the literature (Adler, Ewing, & Taylor, 2008). They have been observed to have weak correlations with citations at the publication level and they are not well representative of individual article impact (Seglen, 1997; Larivière et al, 2016) and can be influenced by highly cited publications (Seglen, 1992). As a reaction to this, some initiatives such as DORA[3] and the Leiden manifesto[4] have warned against the misuse of journal-based indicators in the evaluation of publications and individuals.

On the other hand, high journal impact indicators may indicate a higher probability that some publications in the journal will attract large numbers of citations (although we do not know beforehand which ones will be the most highly cited). In addition, authors tend to see publication in high impact journals as a strong performance in itself since these journals are often highly selective. For instance, Biomedical researchers in the Netherlands perceive the quality and novelty of papers by the impact of journal (namely JIF) in which these papers are published (Rushforth, & Rijcke, 2015). The combined use of journal and publication level impact indicators (so called "composite indicator") has been proposed for evaluating recent publications (Levitt & Thelwall, 2011; Stern, 2014). It has also been shown that using geometric vs. arithmetic mean in calculation of journal impact factor helps to reduce the influence of highly cited publications on its correlation with individual publications (Thelwall & Fairclough, 2015). In a similar line, using journal level metrics in evaluating research has been seen as a relevant practice in some countries (e.g. in Spain, Jiménez-Contreras et al, 2003). In addition, there have been discussions about the potential relevance of journal-based indicators as tools for the analysis of researchers and, particularly, for the potential filtering and selection of academic papers for reading (cf. Waltman, 2016). In this paper we follow up on the latter argument (i.e. the relevance of using journal-based indicators to filter highly cited publications) in contrast to Mendeley readership.

*Justification and aim of this study*

It will be clear that if alternative metrics do not improve the ability of filtering highly cited publications of journal indicators, they don't really pose a true advantage over currently existing measures of impact (e.g. the Journal Impact Factor) for this purpose. The study of the ability of altmetric indicators to identify highly cited publications in comparison with journal-based indicators has shown that journal-based indicators have both a stronger correlation with citations as well as stronger filtering power to identify highly cited publications than for example F1000 recommendations, tweets, blogs as well as other altmetric indicators (Waltman & Costas, 2014; Costas, Zahedi, & Wouters, 2015a). These results reinforce the idea that the above mentioned altmetric indicators do not introduce any advantage over journal-based indicators to identify highly cited publications. Mendeley hasn't been thoroughly studied yet from this perspective. If Mendeley would offer a better filtering solution for identifying highly cited publications than journal indicators, it could be argued that "at least" Mendeley readership represent a true alternative to journal indicators when screening for relevant publications. Actually, in a preliminary study it has been reported that readership scores are more effective at identifying highly cited publications than journal citation scores for the 2011 Web of Science publications (Zahedi, Costas & Wouters, 2015). Thus, as mentioned before, in contrast to other altmetric indicators, this finding indicated for the first time that Mendeley readership scores could represent a valuable tool as an alternative to journal indicators for a more effective filtering of highly cited publications. Due to the relevance of such a result, and since the previous study was limited to only one publication year, in this study we aim to extensively test whether this pattern is also present in data sets with longer publication and citation windows as well as from different scientific disciplines. Thus, the main aim of this paper is to explore the relationship between Mendeley readership and journal citation scores, particularly focusing on whether Mendeley readership scores are able to identify highly cited publications more effectively than journal-based impact indicators.

**Data and Methodology**

This study is based on a dataset of 9,152,360 (77.5%)[5] Web of Science (WoS) publications (articles and reviews) with Digital Object Identifiers (DOI) from the years 2004-2013. The readership data from Mendeley were extracted via Mendeley REST API on February 9, 2015. 86.5% (7,917,494) of all papers have at least one Mendeley readership while 13.5% (1,234,866) of them don't have any. A variable citation window (i.e. citations from 2004 until the end of the year 2014) has been considered for calculating the citation scores. Self-citations have been included in the citation scores in order to keep the same approach for citation and readership data, since it is not possible to calculate something like "self-readership" in Mendeley. Moreover, due to the lack of information on the date of documents added to users' libraries in Mendeley (the date of readership), it is not possible to exactly establish the same citation and readership windows. Hence, we consider the sum of all readership data until 9 February 2015 as the total readership score. The journal citation score and top 10% most highly cited publications[6] in the period 2004-2014 have been calculated for each publication. Only document types 'article' and 'review' were considered. The fields to which publications belong were determined according to the five major fields of science in the 2013 Leiden Ranking classification[7]. The following indicators have been calculated for the different analysis using the CWTS in-house database:

*P:* total number of publications (articles and reviews).

***Total Citation Score (TCS)****: sum of all the citation scores received by the publications in the period of 2004-2014.*

***Total Readership Score (TRS)****: sum of all Mendeley readership scores (RS) received by the publications until February 2015.*

***Mean Citation Score (MCS):*** *average number of WoS citation scores per publication.*

***Mean Readership Score (MRS):*** *average number of Mendeley readership scores per publication.*

***Journal Citation Score (JCS)****[8]: average number of WoS citations received by all publications in a journal in a period of 2004-2014.*

The distribution of the above indicators over time and across their subject fields has been investigated. This has been done in order to provide a general overview of the data and to identify any relevant pattern regarding the density of readership in comparison to citation scores across fields and publication years. A precision-recall analysis (Harman, 2011) has been performed in order to evaluate the ability of readership scores and journal citation scores to identify highly cited publications. In information retrieval, precision is the proportion of retrieved documents that are relevant, while recall is the proportion of relevant documents that are retrieved. Accordingly, in this study for a given selection of publications, precision is the ratio (%) of highly cited publications divided by the total number of publications in the selection, and recall is the ratio (%) of highly cited publications in the selection divided by the total number of highly cited publications (Waltman & Costas, 2014). All the 'top 10%' highly cited publications in the sample have been identified. Then, publications have been ranked by their individual readership scores in a descending order (ties have been sorted randomly) and the precision-recall analysis has been performed. The same process was performed using the journal citation score of the journal of each individual publication. Thus two precision-recall analyses have been produced, one for the readership scores and another one for the journal citation scores. Finally, the values have been plotted, where the *x* axis represents the 'Recall' and *y* axis represents the 'Precision' values. The precision-recall analysis has been done both across publication years (from 2003-2014) and also across subject fields (based on the 5 major fields of science in the 2013 Leiden Ranking classification) of the publications. The precision-recall curves provide visual representations of how precision values correspond with their recall values.

**Results**

*General distribution of citation and readership scores over time*

Table 1 shows the descriptive statistics for the entire publication set used in this study. In general, the average number of citation per publication (MCS) is higher than the average number of readership score per publication (MRS), which means that on average all publications received more WoS citation sores than Mendeley readership scores. The table also shows that the coverage of publications with at least one Mendeley readership is increasing from 2004 to 2009 with a decrease for the most recent years (from 2010 until 2013).

**Table1. General distributions of MRS and MCS indicators of the WoS publications across publication years 2004-2013**

| Pub year | P | Cov | % | TRS | MRS | TCS | MCS |
|---|---|---|---|---|---|---|---|
| **All years** | 9,152,360 | 7,917,494 | 86.51 | 102,051,962 | 11.15 | 132,246,959 | 14.44 |
| **2004** | 540,924 | 458,114 | 84.69 | 6,129,245 | 11.33 | 15,724,035 | 29.07 |
| **2005** | 618,976 | 531,409 | 85.85 | 7,452,051 | 12.04 | 16,706,508 | 26.99 |
| **2006** | 713,864 | 615,637 | 86.24 | 8,697,103 | 12.18 | 16,990,568 | 23.80 |
| **2007** | 788,533 | 682,704 | 86.58 | 9,801,854 | 12.43 | 16,669,281 | 21.14 |
| **2008** | 872,572 | 768,813 | 88.11 | 11,252,702 | 12.90 | 16,084,499 | 18.43 |
| **2009** | 962,262 | 857,585 | 89.12 | 12,547,495 | 13.04 | 15,106,704 | 15.70 |
| **2010** | 1,026,541 | 913,414 | 88.98 | 13,260,840 | 12.92 | 13,026,893 | 12.69 |
| **2011** | 1,120,212 | 987,479 | 88.15 | 12,909,807 | 11.52 | 10,504,765 | 9.38 |
| **2012** | 1,206,707 | 1,030,886 | 85.43 | 11,217,458 | 9.30 | 7,499,214 | 6.21 |
| **2013** | 1,301,769 | 1,071,453 | 82.31 | 8,783,407 | 6.75 | 3,934,492 | 3.02 |

Number of publications (P); Coverage (n. pubs) in Mendeley per publication year (Cov); Total Readership Score (TRS); Mean Readership Score (MRS); Total Citation Score (TCS); Mean Citation Score (MCS)

According to figure 1, MCS steadily decreases over these 10 years; while, on the other hand, MRS first follows a relatively stable pattern with a small increase from 2004 to 2009 and then shows a decrease from the year 2010 onwards, in which MRS is higher than MCS. The higher density of MRS over MCS for publications has also been observed in previous studies on Mendeley (Haustein & Larivière, 2014; Thelwall, 2015; Maflahi, & Thelwall, 2016; Zahedi, Costas, & Wouters, 2015; Costas, Zahedi, & Wouters, 2015a). This suggests that the more recent publications received on average more readership than citation scores. These results support the idea of a faster accumulation of Mendeley readership scores over publications in contrast to citation scores.

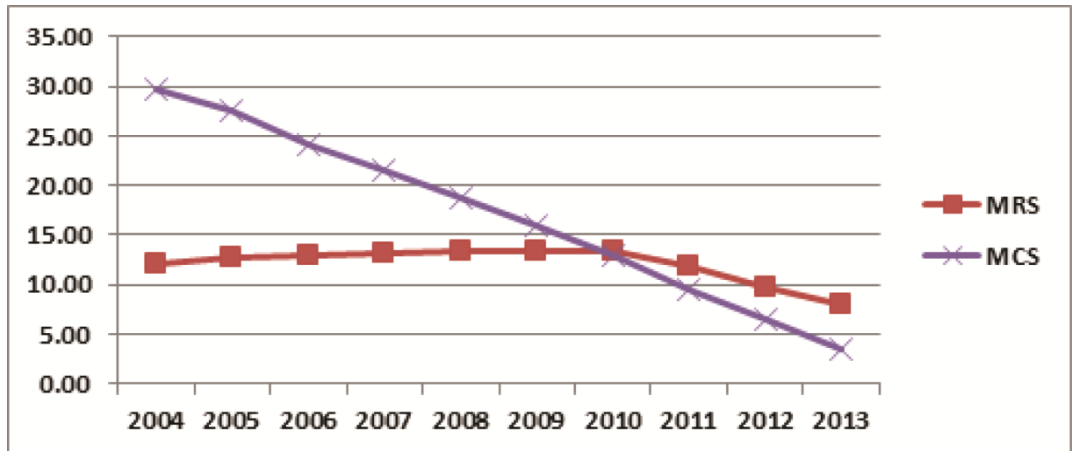

**Figure1. Distributions of MRS and MCS indicators for the WoS publications overtime**
**(x axis shows the publication years and y axis shows the mean scores of citation and readership)**

*General distribution of MCS and MRS indicators across fields*

MRS and MCS indicators have been calculated for the publications based on their main disciplines in the 2013 Leiden Ranking (LR) classification. Table 2 presents the values of MCS and MRS for the 5 major LR fields of science. 'Biomedical and health sciences' is the biggest field with around 36% of all Mendeley-covered publications while 'Social sciences and humanities' is the smallest one in the dataset (7.6%). In terms of coverage of publications in Mendeley (i.e. based on publications (articles and reviews) with DOI), 93% of publications from 'Life & earth sciences' and 92% from 'Social sciences & humanities' have at least one reader in Mendeley, while just 77% of publications from 'Mathematics & computer science' have some readership in Mendeley. Also, the coverage of publications per LR fields with presence in Mendeley increases from 2004 to 2010 with a small decrease for the recent years (from 2011-2013) (see appendix 1 for all publication years). In terms of citation and readership frequency, 'Life and earth sciences' have on average the highest mean readership scores (MRS=18.64) followed by 'Social sciences and humanities' (MRS=18.14), whereas 'Biomedical and health sciences' have the highest mean citation scores (MCS=20.18). Publications from 'Mathematics and computer sciences' exhibit the smallest values both in terms of readership and citation scores (MRS=7.52 and MCS=8.0). In terms of citation and readership density, publications from Social sciences fields have a higher density of readership over citation scores. In contrast, publications from 'Biomedical and health sciences', although with the highest coverage in the sample, exhibit a lower readership density as compared to their citation density. These results are in line with previous analyses (Thelwall, 2015; Costas, Zahedi, & Wouters, 2015b) and indicate that, similar to citation, the readership density of publications varies per fields. Furthermore, large differences in the WoS database coverage across disciplines could affect the density of citations across subject fields, the same holds for the Mendeley database.

Table2. General distributions of MRS and MCS indicators for the WoS publications across LR fields

| Main fields | P | % | Cov | % | TRS | MRS | TCS | MCS |
|---|---|---|---|---|---|---|---|---|
| Biomedical & health sciences | 3,340,837 | 35.90 | 3,033,467 | 90.80 | 45,468,376 | 13.60 | 67,437,722 | 20.18 |
| Life & earth sciences | 1,512,173 | 16.25 | 1,407,153 | 93.06 | 28,189,119 | 18.64 | 26,668,168 | 17.63 |
| Mathematics & computer science | 859,363 | 9.24 | 660,908 | 76.91 | 6,470,579 | 7.52 | 6,877,035 | 8.00 |
| Natural sciences & engineering | 2,878,982 | 30.94 | 2,409,731 | 83.70 | 23,641,874 | 8.21 | 43,656,107 | 15.16 |
| Social sciences & humanities | 714,142 | 7.67 | 659,754 | 92.38 | 12,956,645 | 18.14 | 7,346,205 | 10.28 |

Number of publications (P); Coverage (n. pubs) in Mendeley (Cov); Total Readership Score (TRS); Mean Readership Score (MRS); Total Citation Score (TCS); Mean Citation Score (MCS)

Comparing the distribution of citation and readership scores across fields of science, Figure 2 shows that for fields such as 'Social sciences and humanities' and 'Life and earth sciences' MRS values are higher than MCS values. These are also the fields with the highest coverage in Mendeley. This higher density of readership over citation is even bigger in the field of 'Social sciences and humanities' (MRS=18.14 vs. MCS=10.28). There are also variations in density of MRS vs. MCS by the different LR fields across the different publication years (see figures in appendix 1.2). Basically, for the oldest papers of all disciplines, MCS values are higher than MRS values, while MRS values are higher than MCS in all cases for the most recent years. The case of 'Social sciences and humanities' is different, as MRS outperforms MCS for all years except for the first year 2004 (see figures in Appendix 1.1 & 1.2) indicating that readership scores in this field have a much stronger density as compared to citations over a longer period of time. In order to further explore which subfields within the 'Social sciences and humanities' exhibit higher readership vs. citation densities, MRC and MCS values have been calculated for the individual WoS subject categories (Appendix 2.1). The results show that publications from fields such as Business, Psychology, Sociology, Social and behavioral sciences, Anthropology, Education and educational research and Linguistics are among the fields that have a higher readership density than citation density. Fields such as Chemistry, Oncology, Hematology, Physics, Medicine and Virology have the highest MCS values over MRS (see Appendixes 2.1 & 2.2). These results confirm the idea of important disciplinary differences in readership practices (see Thelwall & Sud, 2015; Costas, Zahedi, & Wouters, 2015b) in a very similar way as it has been observed for citation practices (see Waltman & Van Eck, 2013; Crespo, Li, & Ruiz–Castillo, 2013; Crespo et al., 2014). These differences highlight both different citing and reading practices across fields as well as the disciplinary differences in the coverage of citation and readership databases. Disciplinary differences have also been seen in the use of other academic social networking sites and other online reference managers. For example, Academia.edu is mostly used by academics from Social sciences and humanities in contrast to researchers from physical, health and life sciences, biology, medicine and material sciences with very low usage of this platform (Thelwall & Kousha, 2014; Mas-Bleda, et al., 2014; Ortega, 2015). Similarly, CiteULike is known to be more popular among users from the biomedical domain (Hauff & Houben, 2011). Twitter has been shown to have a good coverage within the field of biomedicine (Haustein et al., 2014a). Twitter is also used by researchers from diverse disciplines such as biochemistry, astrophysics, chemoinformatics (field related to the use of computer techniques in chemistry) and digital humanities, and for different purposes such as scholarly communication, discussions, sharing links (e.g. in fields like economics, sociology and history of science) (Holmberg & Thelwall, 2014).

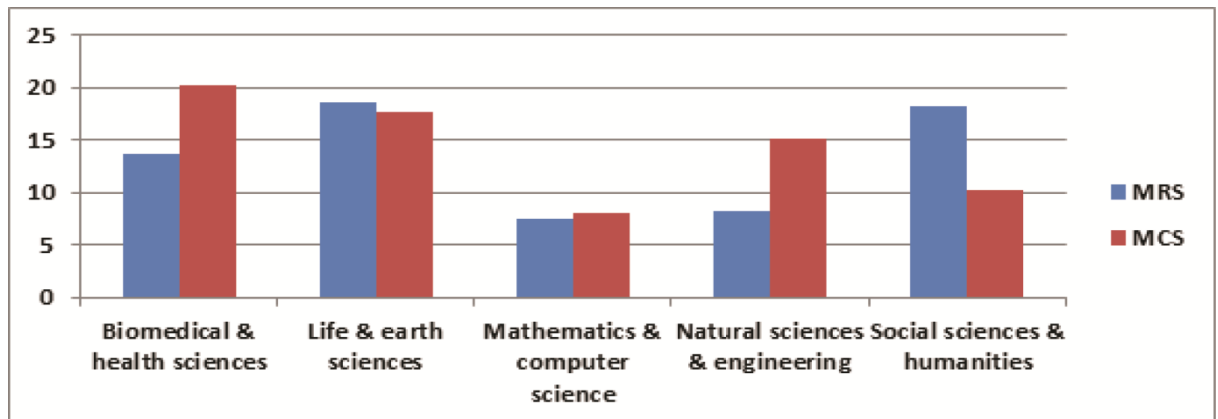

Figure2. Distribution of MRS and MCS indicators for the WoS publications across LR fields
(*x* axis shows the fields and *y* axis shows the mean readership and mean citation scores)

Another study has observed the same variation between fields in the amount of citation and readership scores concluding that in some fields such as Ecology, Evolution, and Behavior and Systematics (based on Scopus subject categories), Mendeley scores are much higher than citations. Also, correlations between these Mendeley readership and citations have been found to have a decreasing trend for recent publications (2011 to 2014) (Thelwall & Sud, 2015).

All in all, the coverage, language and any other biases related to the citation and readership databases could cause important limitations on research assessment and impact indicators, particularly in some fields with low coverage such as social sciences and humanities (Van Leeuwen et al, 2001). For instance, in the humanities, different information behaviours, dependency on print vs. online materials and database's low coverage of non-English publications influence the analysis of scholarly materials (Collins et al., 2012; Hammarfelt, 2014). As an alternative solution to any bias that a database may have, the combined use of citation databases has been proposed (Meho & Sugimoto, 2009). Further research should therefore focus on considering other databases and test if the elements discussed here also hold for them. For now, we still consider that an analysis based on the Web of Science has a strong relevance as this is one of the most common and used data sources for Scientometric and altmetric research.

*Precision-Recall analysis of all publications in the sample*

In order to test which of the two indicators (i.e. Mendeley readership scores or journal citation scores) is more effective to identify highly cited publications, precision-recall analyses have been performed across publication years and subject fields separately. Figure 3 shows the results of the general precision-recall analysis of RS over JCS for all the publications in the dataset over time. According to this figure, the RS (green line) performs better than JCS (blue line) in the whole spectrum of precision-recall in identifying the top 10% most cited publications in all publication years. The figure indicates that, for example, a recall of 0.5 (50%) corresponds with a precision of 0.45 (45%) for RS and a precision of 0.25 (25%) for JCS in the years 2004-2013. This means that if we want to select half of all highly cited publications in the dataset in each year, we have an error rate of 55% when the selection is made based on readership scores, and an error rate of 75% when the selection is made based on journal citation scores. Actually, error rate refers to the share of highly cited papers that cannot be identified by one of these two indicators (RS or JCS). In the precision-recall figure, by drawing a vertical line from the recall axis for example from the recall point of 0.5 (50%) crossing the RS and JCS lines, and drawing a horizontal line from there to the precision axis, it shows that the recall of 50% corresponds to a precision levels of 45% for RS and of 25% for JCS. This means that the error rates for RS is 100-45=55% and for JCS is 100-25=75%. The results of the figures are straightforward; the green line always outperforms the blue line in terms of precision in the whole spectrum of recall. Hence we can conclude that readership scores identify highly cited publications better than journal citation scores for all the publication years in our dataset. This is a very important result as it has not been observed before for other altmetric sources (cf. Costas, Zahedi, & Wouters, 2015a; Waltman & Costas, 2014).

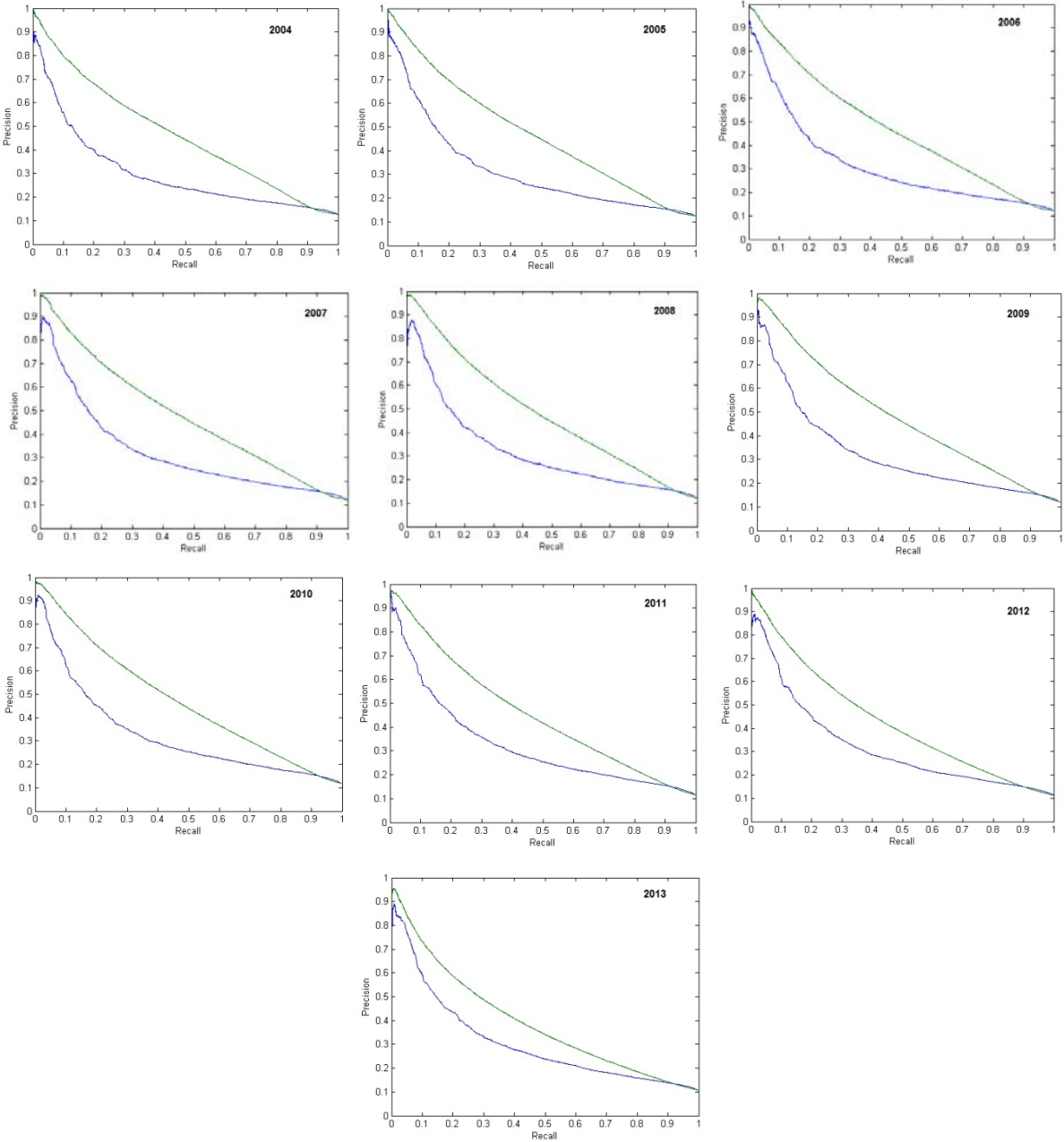

**Figure3. General Precision-recall curves for JCS (blue line) and RS (green line) for identifying top10% most highly cited WoS publications from the years 2003-2014 left to right (x axis represents the 'Recall' and y axis represents the 'Precision' values)**

*Precision-recall analysis of publications across their disciplines*

In this section, the precision-recall analysis has been performed across disciplines. Results indicate that RS also outperforms JCS in identifying highly cited publications for all major fields of science. All the figures are similar, essentially resembling the general patterns in Figure 3. These results are in line with the result obtained for the 2011 WoS publications (Zahedi, Costas, & Wouters, 2015) confirming the better capacity of RS over JCS in identifying highly cited WoS publications for fields of science. Thus, this pattern can be considered to be robust both across disciplines and years (see also appendix 3). The only noticeable exception is the field of 'Mathematics & computer science'. In this field, JCS outperforms RS both in the lower (below 10%) and higher (above 80%) levels of recall. For example, for the publications from the years 2004 to 2009, RS outperforms JCS until the recall point of 0.5 (50%) while, for the most recent years (from 2010 onwards), there is a small advantage of JCS over RS particularly from the recall point of 0.5 onwards. A potential explanation for this exception is that this is the field with the lowest coverage of publications saved in Mendeley (76.91%) of the publications in this field are covered by Mendeley) as well as the field with the lowest density of both citation and readership scores compared to the other fields in the study. These lower coverage and density values could be more easily affected by all kinds of random effects coming from citation and also readership processes[9] (cf. Waltman, Van Eck, & Wouters, 2013), thus having a greater influence on the patterns observed for this discipline. According to the literature, the low citation rates of Mathematics and computer science compared to fields such as Chemistry or Physics can be also related to the specific publication and citation behaviours in these fields (Korevaar & Moed, 1996; Seglen 1997). For instance, scholars from fields like Mathematics and computer science are known to publish more in formats such as research reports and conference papers which are not included in citation databases such as Web of Science (Moed et al., 1985; Bornmann et al., 2008). Also, Mathematics is a discipline with a relatively low number of references per paper as compared to other disciplines (Vieira & Gomes, 2010; Glänzel & Schoepflin, 1999). This lower level of references per paper may explain the lower density of citations per paper in the field (i.e. there are fewer references (citations) pointing to other Mathematics papers) as well as lower numbers of Mendeley readership (i.e. Mendeley users from Mathematics would save fewer records in their Mendeley libraries). Other reasons for the low rates of readership also include the different orientation, uptake and use of Mendeley among scholars in this field. Users from Mathematics and computer science seem to be more oriented towards other reference managers such as BibSonomy (Hauff & Houben, 2011), which may support the idea that Mendeley is not the most popular online reference manager tools among the users of these fields.

All in all, the results of the precision-recall analysis highlight the importance and potential of Mendeley readership as a tool for research evaluation. This suggests that readership data can be used as a relevant tool in finding highly-cited publications. This result together with the fact that Mendeley readership are available both openly and also much earlier than citation as well as their potential in revealing an early impact of publications (Maflahi & Thelwall, 2016), put an emphasize on an additive value that readership data offer in case of its usage beside other impact indicators for any research evaluation and scientometrics purposes.

**Discussions and conclusions**

This study presents a large-scale analysis of the distribution and presence of Mendeley readership scores over time and across disciplines. Precision-recall analysis has been used to test the ability of Mendeley readership scores to identify WoS highly cited publications, particularly in comparison with journal citation scores. Our results show that 86.5% of the publications in our dataset were covered in Mendeley with at least one reader. The coverage of publications with some Mendeley readership increased from 2004 to 2009 with a small decrease from 2010 onwards. Disciplinary differences have been found in terms of both citation and readership density. These differences of readership density could be explained by the different levels of awareness and adoption regarding the use of Mendeley in the scholarly practice of researchers (Ortega, 2015) or by the use of other reference managers such as BibSonomy or CiteULike (Hauff & Houben, 2011) by scholars from different fields. However, further research on this point is still needed.

The main conclusions of this study can be summarized as follows:

   a) *Steady increase of Mendeley readership scores for the earliest publication years and decreasing pattern for the most recent ones.*

The average readership per publication steadily increases from 2004 until 2009, with a small decrease for the most recent years (i.e. 2010 onwards). This pattern is observable for all fields. These results are in line with those of Thelwall & Sud (2015) for a selection of Scopus thematic categories (including agriculture, business, decision science, pharmacy, and the Social sciences) and LIS journals (Maflahi & Thelwall, 2016). These authors found very similar steady increasing patterns for Mendeley readership for older years with a decrease for the most recent ones. A plausible explanation for this pattern (as opposed to the consistently higher average values of citations per paper for the older years) is that citations are events that can happen several times (i.e., a paper can be cited multiple times), but a paper can only be saved once by each Mendeley user. Thus, the maximum number of readership a paper can achieve is the total number of users in Mendeley, while the number of citations a paper can receive has basically no upper bound. Moreover, the removal of papers from Mendeley libraries[10] by users can contribute to explain the patterns observed for the older years. Thus, in order to maintain manageable libraries, Mendeley users could decide to remove the older and less useful publications from their reference managers. As a result, citations would always accumulatively increase over time as publications have more time to be cited, while the number of readership could actually decrease as users would remove older references from their libraries. Moreover, as pointed by Thelwall & Sud (2015), Mendeley was launched and became available in 2008 and consequently became popular afterwards. This may contribute to explain the increase in MRS values from 2008 to 2009.

Another possible reason for the decreasing pattern of readership for recent publications could be the delay between the publication of the paper and the time needed by the users to spot it and decide to save it in their libraries. In other words, the declining pattern for the most recent years is likely indicating some kind of delay in the accumulation of readership for the most recent publications. Finally, variations in the uptake of Mendeley across fields and the increasing popularity of other reference managers in some fields, as well as changes in the preferences of users in their reference manager choices (e.g. preferring Zotero over Mendeley) might have played an influence on the lower counts of Mendeley readership during the most recent years. However, the lack of reliable information on the uptake of reference manager among different types of users make difficult to determine the true importance of such as pattern. In any case, it is important to notice how even with this delay in the accumulation of readership, they accumulate faster than citations during the three most recent years.

*b) Higher density of Mendeley readership scores over citations for the most recent years and most disciplines*

Our results show that the density of Mendeley readership is higher than that of citations for the most recent years and for most of the disciplines. These results suggest the potential advantage of Mendeley readership over citations for the analysis of impact of the most recent publications and particularly in the field of Social sciences, which is also a field that traditionally is not well represented by citation databases (Nederhof, 2006). Thelwall & Sud (2015) suggested that the faster uptake and the stronger density of Mendeley reader counts for the most recent years could be seen as a good proxy for "early scientific impact" for articles from recent years and also for fields with higher levels of Mendeley use. However, our results also show that as time passes and more citations accumulate, they tend to outperform the values of readership (which tend to remain stable) after around 3-4 years, although again this varies across disciplines. For example, the readership advantage over citations lasts longer in the Social sciences than in the Natural sciences (see appendix 2.1). Maflahi & Thelwall (2016) found similar patterns for a set of LIS journals.

These results suggest that Mendeley readership scores can work as an important source to reflect evidence of "early impact" of scientific publications since, as shown in this study as well as in a previous analysis (Thelwall & Sud, 2015), readership occur and are available earlier than citations during the first years after publication. However, more research is necessary in order to better disentangle the true motivations of Mendeley users and differences between citations and Mendeley readership during the first years after publication of the articles.

*c) Higher filtering ability of highly cited papers by Mendeley readership scores in contrast to journal citation scores*

The most important result of this study shows that Mendeley readership data can work as a relevant tool to identify highly cited publications in WoS. This finding is robust both across most major fields of science and publication years. In contrast, other altmetric indicators (e.g. F1000 recommendations, Twitter, blogs, etc.) have not been found to have such a property, particularly in their comparison with journal citation scores as a benchmark tool to identify highly cited publications. Based on these results, it can be concluded that Mendeley readership can indeed play a role as an alternative approach (to journal-based impact indicators) to find highly-cited outputs, being the only one of all altmetric sources exhibiting such possibility.

Although we haven't approached the issue of prediction of later highly cited publications, as it would be necessary to study early readership counts (which are currently not available in the data provided by Mendeley); it could be argued that this good filtering ability of Mendeley readership could be seen also as a strong indication of a potential predictability of future highly cited publications, particularly if we take into account its faster uptake (i.e., Mendeley readership are accumulated earlier than citations). Therefore, as suggested by Thelwall & Sud (2015) "future work with early Mendeley reader counts and later citation counts for the same set of articles is urgently needed to check this hypothesis" of whether Mendeley readership can predict future citations, and in this case, also highly cited papers.

*Final remarks*

The results of this study show that Mendeley readership scores are an effective tool to filter highly cited publications. This result, together with the moderate correlations between citations and readership found in previous studies (Thelwall & Wilson, 2015; Haustein et al., 2014b) as well as the "pre-citation role"[11] that is expected from Mendeley readership (i.e. that Mendeley users save documents in their libraries to cite them later, cf. Haustein, Bowman & Costas, 2015; Thelwall & Sud, 2015) make it possible to argue that Mendeley readership and citations are two different but connected processes that could be capturing a similar type of impact. However, from a more conceptual point of view, saving a document in Mendeley and citing it are two fundamentally different acts (Haustein, Bowman & Costas, 2015). Thus, considering the broad spectrum of reasons why Mendeley users may save documents in their libraries (for example, not only to cite them later, but also to use them for reading, teaching, self-awareness, individual non-academic interests, personal curiosity, etc.), it would not be correct to fully assimilate Mendeley readership impact to citation impact. Mendeley users cannot be expected to adhere to the same norms and expectations when they save a document as when they cite it[12] and clearly more research is necessary in order to better understand the differences and similarities between these two metrics.

Finally, there are also important technical issues (e.g. differences between the bibliographic metadata reported by Mendeley and WoS) that need to be considered and that can influence the data retrieval and the matching of records based on different identifiers (such as DOI, titles, journals, publication years, etc.) and hence can have an influence in the number of readership per publication (Thelwall, 2015; Zahedi, Bowman, & Haustein, 2014).

Although this study emphasizes the ability of Mendeley readership to identify highly cited publications and its role as a potential evaluative tool, more research is necessary to explore the abovementioned issues and limitations as well as to reveal more accurately the meaning of Mendeley readership and its potential value for research evaluation purposes. Follow up research should continue to explore the conceptual meaning of Mendeley readership and its relationship with citation indicators, as well as study whether Mendeley readership can be used to predict future citation. The disciplinary differences in the database coverage on which the citation and readership data are based is an important factor that should be considered when interpreting the results, and further research should focus on determining the potential influence that different levels of coverage may have on the value of Mendeley readership over journal indicators for all disciplines of Science.

**Acknowledgment**

This paper is an extended version of a paper accepted for oral presentation at the 15th International Conference on Scientometrics and Informetrics (ISSI), 29 Jun-4 July, 2015, Bogazici University, Istanbul (Turkey). The authors are grateful to Henri de Winter (CWTS) for his support on the Mendeley data collection and data management for this study. Also, special thanks to Ludo Waltman, Alex Rushforth (CWTS) and the anonymous referees of the journal for their valuable comments on this paper. Zohreh Zahedi was partially funded by the Iranian Ministry of Science, Research, & Technology scholarship program (MSRT grant number 89100156).

**Endnotes**
1. http://blog.mendeley.com/elsevier/mendeley-and-elsevier-2-years-on/
2. Users in Mendeley can view only the historical overview of readership (the last 12 months) of their own documents saved in Mendeley (this information is not yet available via API and for all the documents saved in Mendeley by all users).
3. DORA: San Francisco Declaration on Research Assessment: www.ascb.org/dora/
4. Leiden Manifesto for Research Metrics: www.leidenmanifesto.org/
5. 77.5% of all WoS articles and reviews from the years 2004-2013 have a DOI.
6. Top 10% publications are publications that belong to the top 10% quartile of the most cited publications in their fields (i.e. Web of Science Subject Categories) and publication years. We have followed the methodology by Waltman & Schreiber (2013) for the calculation of percentile based indicators, although in this case proportionally assigned publications to the top 10 percentile have been considered as fully top 10% highly cited publications. Also all articles and reviews in the WoS database (i.e. including papers without DOIs and not covered by Mendeley) are considered for the determination of the top 10% highly cited publications.
7. www.leidenranking.com/ranking/2013
8. JCS calculation is based on all outputs of the journals (i.e., regardless of having or not DOIs and even if not all of them are covered by Mendeley).
9. The citation process is known for being "noisy" and influenced by multiple random factors that limit the relationships between citation and scientific impact (see Waltman, Van Eck, & Wouters, 2013). In a similar manner, we can argue that similar noisy factors can influence the relationship between the act of saving in Mendeley, citations and scientific impact.
10. According to William Gunn (Director of Scholarly Communications in Mendeley), "When a user deletes their account and all their documents, the readership of that document doesn't change, until the batch clustering process is re-run and the new number of metadata records is generated. The same applies when a user deletes a record from their library. In summary, the count of records can increase nearly instantaneously, but only decreases periodically" see:
www.niso.org/apps/group_public/view_comment.php?comment_id=632
www.niso.org/apps/group_public/view_comment.php?comment_id=610
11. Results of a survey on Mendeley showed that 85% of respondents have saved documents in Mendeley to cite them later (Mohammadi, Thelwall, & Kousha, 2015), which would support the idea of Mendeley readership as a "pre-citation" event (cf. Haustein et al., 2015).
12. For instance, Mendeley users don't necessarily follow the Mertonian norms of "communism", "universalism", "disinterestedness" and "organized skepticism" (Merton, 1973) as pointed out by Haustein et al. (2015) when they select a document to be saved in their libraries, while they could be more driven by these norms when selecting a document for citation.